# Study of time resolution by digital methods with a DRS4 module*

DU Cheng-Ming(杜成名) [1,2]　　CHEN Jin-Da(陈金达) [1;1)]　　ZHANG Xiu-Ling(张秀玲)[1]

YANG Hai-Bo(杨海波) [1]　　CHENG Ke(成科) [1,3]　　Kong Jie(孔洁)[1]　　HU Zheng-Guo(胡正国)[1]

SUN Zhi-Yu(孙志宇)[1]　　Su Hong(苏弘)[1]　　XU Hu-Shan(徐瑚珊)[1]

[1] Institute of Modern Physics, Chinese Academy of Sciences, Lanzhou 730000, China
[2] University of Chinese Academy of Sciences, Beijing 100049, China
[3] Northwest Normal University, Lanzhou 730070, China

**Abstact:**

A new Digital Pulse Processing (DPP) module has been developed, based on a domino ring sampler version 4 chip (DRS4), with good time resolution for LaBr$_3$ detectors, and different digital timing analysis methods for processing the raw detector signals are reported. The module, composed of an eight channel DRS4 chip, was used as the readout electronic and acquisition system to process the output signals from XP20D0 Photomultiplier Tubes (PMTs). Two PMTs were coupled with LaBr$_3$ scintillator and placed face to face on both sides of a radioactive positron $^{22}$Na source for 511 keV ɤ-ray tests. By analyzing the raw data acquired by the module, the best coincidence timing resolution is about 194.7 ps (FWHM), obtained by the digital constant fraction discrimination (dCFD) method, which is better than other digital methods and analysis methods based on conventional analog systems which have been tested. The results indicate that it is a promising approach to better localize the positron annihilation in positron emission tomography (PET) with time of flight (TOF), as well as for scintillation timing measurement, such as in TOF-ΔE and TOF-E systems for particle identification, with picosecond accuracy timing measurement. Furthermore, this module is more simple and convenient than other systems.

## 1 Introduction

DPP (digital pulse processing) modules are used to process raw signals from detectors, with characteristics, including the ability to reduce drift, archive data for later analysis and conveniently adjust analysis parameters for better results. All information contained in the original waveform can be acquired directly by digitized signals from the detectors. For example, the signal rise time and maximum amplitude can be easily extracted from the digitized waveform data. At present, various kinds of DPP modules have been realized using switched capacitor circuits, for example MAGIC [1] and MEG [2]. It can also be useful for PET (positron emission tomography) scanners and portable oscilloscopes [3, 4]. DPP-based systems are more simple and convenient than other systems.

The Paul Scherrer Institute, in Switzerland, has designed a new DPP module, the DRS4 chip [5, 6], which is based on a switched capacitor array (SCA) [3-7]. Its characteristics of high channel density, high analog bandwidth of 950 MHz, and low noise of 0.35 mV make this chip ideally suited for high precision waveform digitization [4, 7]. In order to upgrade the read-out circuits of TOF-PET (time-of-flight PET) system, a new type of readout and acquisition system based on the DRS4 chip has been developed by the Nuclear Electronics Group in the Institute of Modern Physics, Chinese Academy of Sciences. This module is adopted to construct the read-out and acquisition system. The module consists of a DRS4 chip, an ADC chip, a FPGA and other devices. It is capable of performing eight channel waveform sampling and digitization at a maximum sampling rate of 5.12 gigabit samples per second (GSPS) with 66 dB input dynamic range [4, 7]. The internal time resolution between channels is about 52 ps as tested by the cable delay method [7]. The input ranges for the system can be adjusted to match different kinds of detector signals. The channels can be daisy-chained for larger sampling depth [7, 8]. In the following text, the readout and acquisition system based on the DRS4 chip will be called the DRS4 module for abbreviation.

*This work was supported by the Science Foundation of the Chinese Academy of Sciences (210340XBO)，National Natural Science Foundation of China (11305233,11205222) ,General Program of National Natural Science Foundation of China (11475234), Specific Fund of National Key Scientific Instrument and Equipment Development Project (2011YQ12009604) and Joint Fund for Research Based on Large-Scale Scientific Facilities (U1532131).
1)  E-mail : chenjinda@impcas.ac.cn

The DRS4 module has been employed with fast scintillators and PMTs for coincidence timing measurements. During the experiments, the arrival time difference of two coincident $^{22}$Na $\gamma$-rays were detected by two LaBr$_3$ scintillators coupled with XP20D0s [4, 9], by means of various DPP techniques, such as leading edge discrimination (LED) [9, 10], digital constant fraction zero-crossing discrimination (CFD zero-crossing) [9-12] and digital constant fraction discrimination (dCFD) [9-11]. Each method was repeated a large number of times to optimize the parameters for the best resolution and accuracy. The conventional analog method with the constant-fraction zero-crossing technique was compared with the DPP techniques.

The differences of the calculated arrival times were plotted as a histogram and fit with a Gaussian distribution. The full-width at half maximum (FWHM) of the fit is used as the resolution.

## 2 Experiment setup

Two LaBr$_3$ scintillators ($\phi 20 \times 5$ mm$^3$) coupled with XP20D0s were placed on opposite sides of a $^{22}$Na source and irradiated with 511 keV $\gamma$-rays from a distance of 15 cm, as shown in Fig. 1. The applied bias voltages between the anode and cathode were increased to 1400 V to get a linear photoelectric signal output. The radioactive $^{22}$Na source was fixed in an aluminum holder with a 2 mm lead collimating aperture.

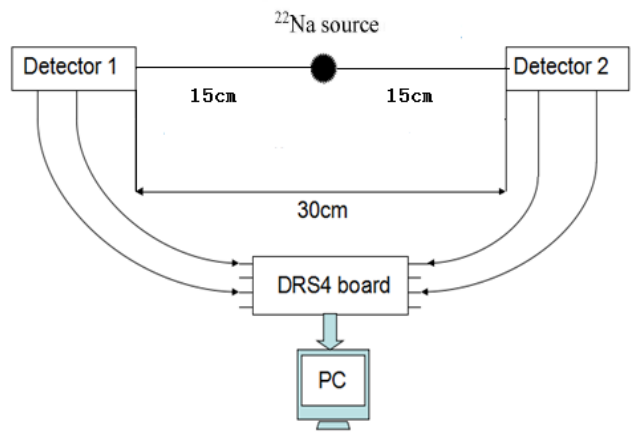

Fig. 1 Schematic diagram of the two LaBr$_3$ detectors and $^{22}$Na source using
the DRS4 module as read-out and data acquisition system.

In this experiment, the DRS4 module worked at a sampling rate of 5.12 GSPS, with an individual channel depth of 1024 cells for each channel, and the sampling time interval was about 200 ps. The full time range of each channel is about 200 ns [5-7]. If the amplitude of the input signal exceeds the thresholds of the DRS4 module in the default time period, the trigger state will be turned on in internal triggering mode. Some period of time behind the trigger, the input signals will be sampled and stored in a series of capacitors at high sample rates under the control of a shift register and digitally converted to discrete waveform data with a commercial ADC operating at lower sample rate [7]. Then these data will be used to process for energy and time information by using digital methods. In the measurement, the dynode signals from the two detectors are sent to channel 0 and channel 1 of the DRS4 module, while the anode pulses are sent to channel 2 and channel 3. The shapes of the dynode and anode signals from the two LaBr$_3$ detectors for 511 keV γ rays acquired by the DRS4 module are shown in Fig. 2. Ten thousands of events (four signals) were sampled and transformed to digital data by the DRS4 module in the experiment.

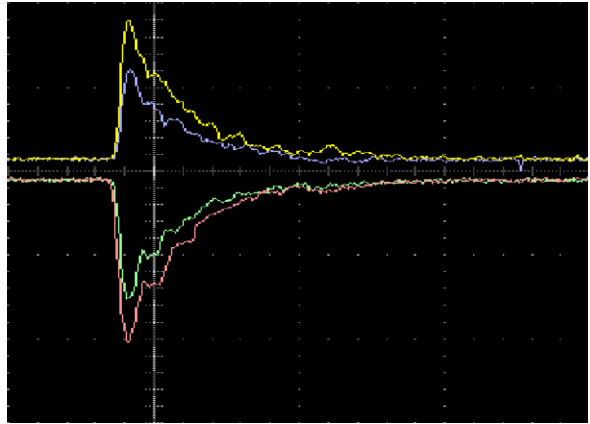

Fig. 2 Dynode (positive) and anode (negative) signals are shown on the interface of the DRS4 module.

The DC offset of each sampling cell needs to be calibrated and removed from the sampling results, and as the parameters of the transistors in the chip normally change gradually, timing calibration is also necessary. The calibration methods for time and amplitude are described in Ref. [7].

During the experiment, an analog system constructed by traditional NIM electronic modules and a CAMAC system with the constant-fraction zero-crossing technique is compared with the DRS4 module. The schematic diagram of the analog system is shown in Fig. 3.

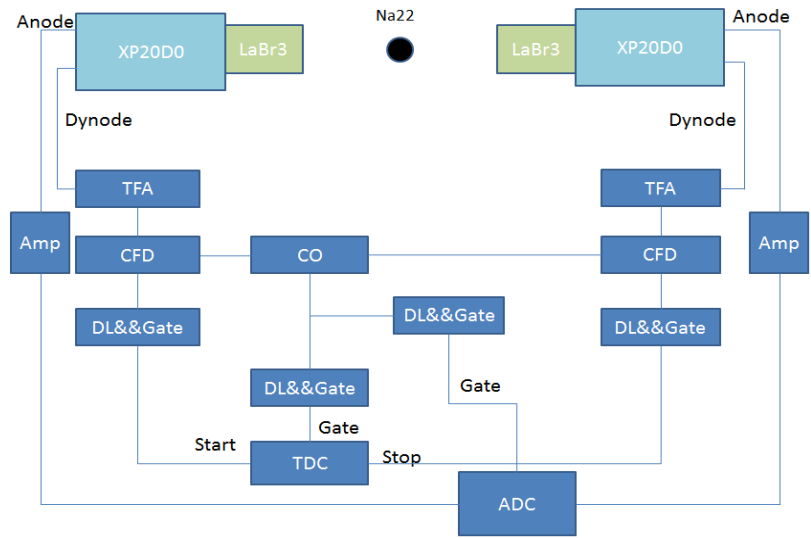

Fig. 3 Schematic diagram of the analog system

The details of the electronic modules are listed as follows: amplifier (Amp): ORTEC 572 for amplifying anode signals; timing filter amp (TFA): ORTEC 474 for filtering and amplifying dynode signals; constant fraction discriminator (CFD): ORTEC CF8000 for signal discrimination; coincidence (CO): CO4020 for selecting coincidence events; delay and gate generator (DL&&Gate): ORTEC GG8000 for generating start ,stop and gate (100ns width) signals to time-digital-converter (TDC: Phillips 7187) and gate (2 us width) to amplitude-digital-converter (ADC: Phillips 7164) with accurate timing.

## 3 Digital analysis methods and Results

### 3.1 Waveform Reconstruction

The raw data of waveform sampling are stored as the following forms: the data arrays of amplitude [1024] and time [1024]. The original 2-byte integer amplitude voltage values and the 4-byte floating point effective time bin widths are converted to ROOT data by the decoding rules [8]. Fig. 4 illustrates the reconstructed energy signal (a) and time signal (b) from the $LaBr_3$ detectors for 511 keV γ-rays. These signals have typical time characteristics like fast rise-time and short decay time for the detectors. The first 200 samples of each waveform should be averaged and subtracted from the rest of the waveform to compensate for the DC offset [11].

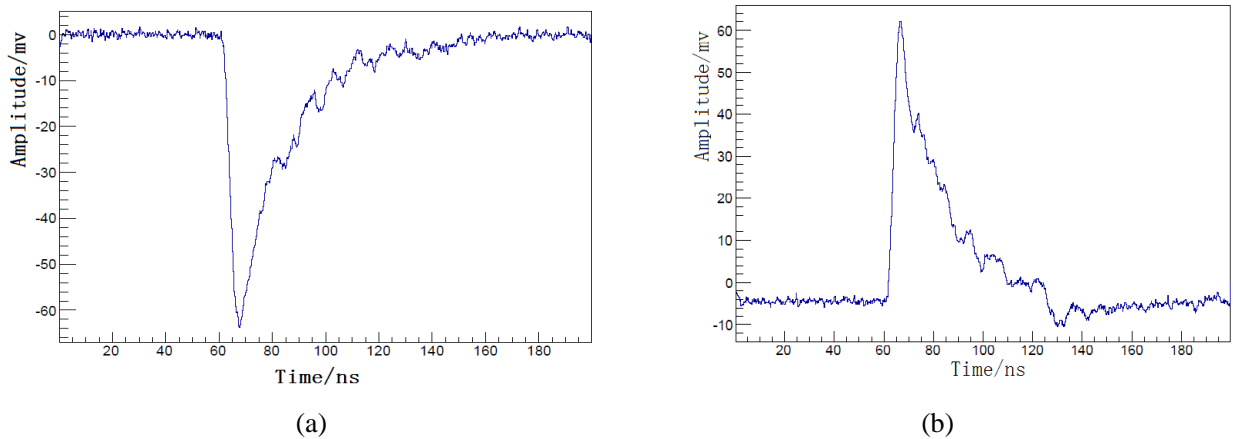

(a)                                      (b)

Fig.4 The reconstructed energy signal (anode-negative pulse) and time signal (dynode-positive pulse).

### 3.2 Energy and amplitude discriminator

The total energy of the measured pulse is proportional to the sum of all created scintillation photons. The pulse area between the digitized waveform and the baseline is related to γ ray energy. The value of the pulse area is simply to sum the amplitude value of each channel over the whole waveform range by a digital integration method [4, 9]. The obtained energy resolutions by the DRS4 module at 511 keV are 3.42% and 4.10% respectively for the two detectors. The energy distributions obtained by the analog system, by contrast, have the energy resolutions at 511 keV of 4.25% and 4.60% at FWHM. The γ ray energy distributions of the same detector obtained by DRS4 module and analog system are shown in Fig. 5. All the photo-peaks of the energy spectrum were normalized to 511 keV. The results indicate that all the photo-peaks can be clearly separated from the Compton scattering and the DRS4 module is better than the analog system for 511 keV γ-ray energy measurement.

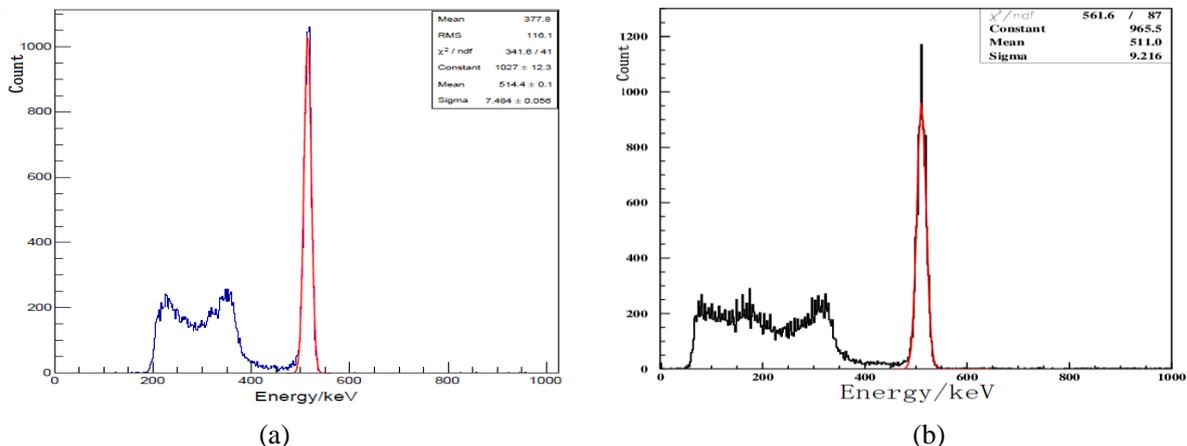

(a)                                      (b)

Fig.5 The γ ray energy distributions obtained by DRS4 module (a) and analog system (b) with $^{22}$Na source.
The energy resolutions are 3.42% (a) and 4.25% (b).

The signals with energy below 400 keV are attributed to Compton scattered photons, as shown in Fig. 6. The amplitude distribution of time signals from the LaBr$_3$ detector obtained by the DRS4 module is illustrated in Fig. 7. The energy signals in the energy window (490, 530) keV and time signals in the amplitude window (75, 95) mV were simultaneously chosen to discriminate the appropriate events. Then these events were used to determine the time information by using DPP methods subsequently.

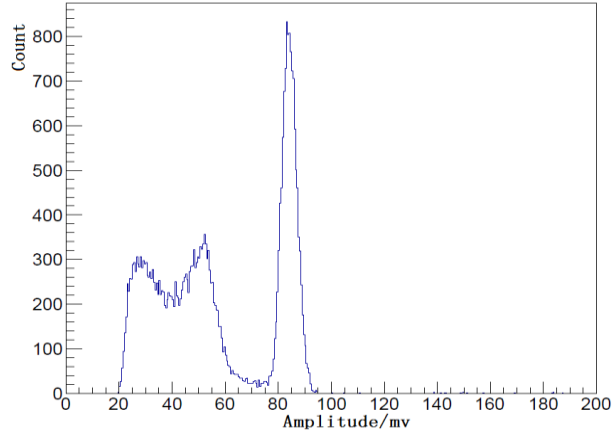

Fig. 7 The amplitude distribution of time signals from the LaBr$_3$ detector
obtained by the DRS4 module with $^{22}$Na source for γ ray test.

### 3.3 Digital methods for timing measurements

Various digital time analysis methods are compared with the dCFD method to obtain the time difference distribution and time resolution.

#### 3.3.1 dCFD method

In this paper the signal processing by the dCFD method is as follows: First, it is necessary to confirm the maximum digitized value (MDV) of the amplitude of the pulse and then set a threshold (Vth) based on a predefined fraction (p) of the MDV [11]. Then, two adjacent sample points (a1, t1), (a2, t2) near the threshold level were chosen for each signal, with the amplitudes in the range of a1<threshold<a2. Meanwhile, a linear interpolation [7, 8] was performed between (a1, t1) and (a2, t2) to estimate the actual zero-crossing time T1(T2) for each time pulse as illustrated in Fig. 8 (a). In the experiments, the threshold was set equal to 6% of the MDV to obtain the distributions of time difference (T1-T2) for the two LaBr$_3$ detector signals. The distribution of the arrival time differences is shown in Fig. 8 (b). The best time resolution is 194.7 ps.

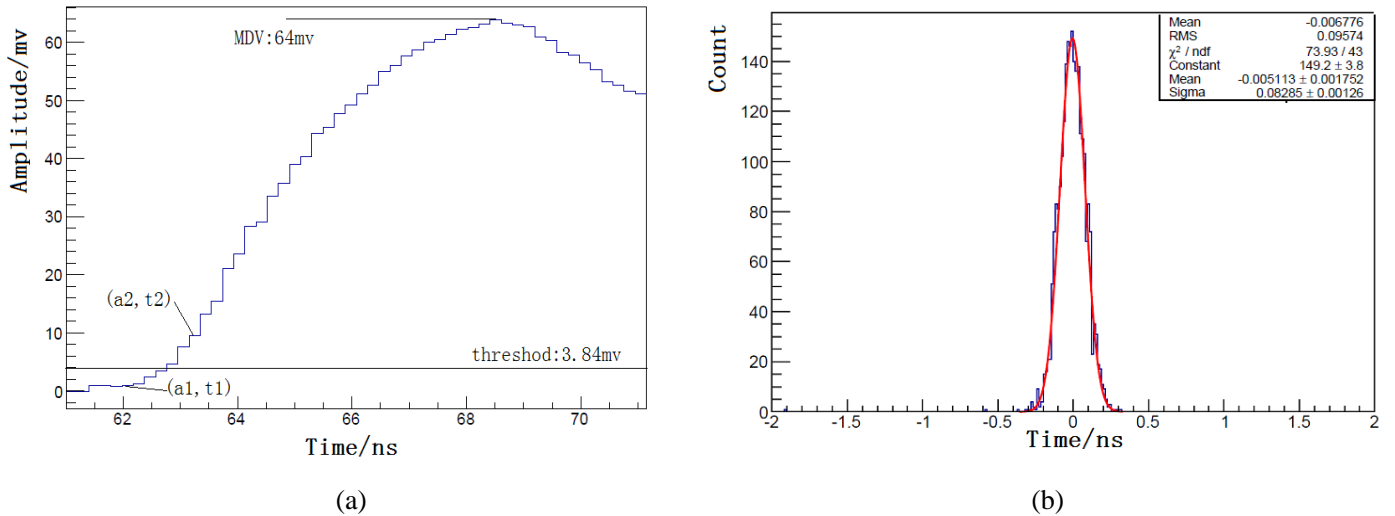

(a)                                          (b)

Fig. 8 Sketch of the linear interpolation to estimate the actual zero-crossing time T1 for dCFD method (a).
The best time resolution is 194.7 ps obtained by the dCFD method (b).

3.3.2 LED method

The LED method is similar to the dCFD method, with a constant threshold (6% of the mean values of pulse amplitudes in the 511 keV peaks area) in the experiment [9, 11]. The best time resolution obtained by the LED and CFD zero-crossing method was 210.9 ps.

3.3.3 CFD zero-crossing method

The CFD zero-crossing method is the digital version of the analog CFD method. In this process, the original signal is delayed from 1 to 5 time steps, amplified and inverted at first. Then the processed signal is added to the original signal. After this process, the unipolar signal is optimized and transformed into a bipolar pulse. The bipolar pulse crosses the time-axis at zero amplitude, with the crossing-zero time at a constant fraction of the height of the original pulse [11, 12]. Two adjacent sample points (a1, t1), (a2, t2) near the zero amplitude level are chosen for each signal, with the amplitudes in the range of a1<0<a2. Meanwhile, a linear interpolation is performed between (a1, t1) and (a2, t2) to estimate the actual zero-crossing time T1 (T2) for each time pulse. The best time resolution obtained by the CFD zero-crossing method is 229.3 ps.

3.3.4 Comparison of different digital timing methods

The time resolution obtained by the dCFD method is better than the LED, CFD zero-crossing and analog methods. It is also better than other methods described in the literature, such as the mean PMT pulse model [9] and Gaussian rising edge fitting method [4], as shown in Table 1. Considering the results and also the precision and the computer processing speed (judged by the operation times of algorithms), the dCFD method was finally adopted for timing measurement in this paper.

Table 1. Time resolution obtained by different DPP methods

| Methods | Time resolution/ps |
|---|---|
| dCFD | 194.7 |
| LED | 210.9 |
| CFD zero-crossing | 229.3 |
| Mean PMT pulse | 195.4 [9] |
| Gaussian rising edge fitting | 206 [4] |
| analog | 254.7 |

**3.4 Optimizing parameters for the dCFD method**

The constant fraction discrimination (CFD) method analyses a large number of the leading-edge-timing measurement results and optimizes the timing moment to obtain the best time resolution on the basis of the following rules. The timing moment is chosen to be that with the minimum time jitter when the pulse crosses a certain constant fraction level of the full amplitude. We investigated the influencing parameters [11] of the time resolution with the dCFD method, including the fitting range and the constant-fraction factor p, which was equal to $V_{th}$/MDV in the experiment. In order to optimize the resolution and the accuracy, the numbers of points used in the analysis before and after $V_{th}$ were varied, which were referred to as pre-$V_{th}$ and post-$V_{th}$, respectively.

The relationship between p and the time resolution is illustrated in Table 2 and the relationship between the fitting range and time resolution is shown in Table 3. Table 2 shows that the change of constant-fraction factor from 3% to 10% does not make much difference for time resolution. The best time resolution obtained is 194.7 ps with the threshold of MDV set to be 6% and the fitting range set to be pre-Vth=4 and post-Vth=3.

Table 2. The relationship of the Constant-fraction factor (P) and the time resolution.

| P | Time resolution/ps | Error/ps |
|---|---|---|
| 2% | 212.8 | 3.08 |
| 3% | 199.9 | 2.89 |
| 5% | 198.5 | 2.87 |
| 6% | 194.7 | 2.96 |
| 7% | 198.0 | 3.08 |
| 10% | 200.6 | 2.90 |
| 15% | 202.5 | 2.98 |
| 25% | 210.6 | 3.24 |

Table 3. The relationship of the fitting range and the time resolution.

| Fitting ranger | Time resolution/ps | Error/ps |
|---|---|---|
| Pre-$V_{th}$=5 Post-$V_{th}$=5 | 200.69 | 3.06 |
| Pre-$V_{th}$=4 Post-$V_{th}$=4 | 195.12 | 2.73 |
| Pre-$V_{th}$=4 Post-$V_{th}$=3 | 194.7 | 2.72 |
| Pre-$V_{th}$=4 Post-$V_{th}$=2 | 197.14 | 2.84 |
| Pre-$V_{th}$=4 Post-$V_{th}$=1 | 199.38 | 2.89 |
| Pre-$V_{th}$=2 Post-$V_{th}$=3 | 201.49 | 3.08 |
| Pre-$V_{th}$=3 Post-$V_{th}$=3 | 197.47 | 2.87 |
| Pre-$V_{th}$=3 Post-$V_{th}$=2 | 198.36 | 2.89 |

## 4 Discussion and Conclusion

A pair of LaBr$_3$ scintillator detectors and a new flexible waveform sampling readout electronics and acquisition system, based on a DRS4 chip, were realized and tested in this experiment. The ability of different DPP methods for testing and processing the arrival time difference of a pair of digitized radiation detectors signals has been examined .The best time resolution acquired with the DRS4 module is 194.7 ps at FWHM, which is obtained by the dCFD method with a constant-fraction factor of 6% and a fitting range of Pre-$V_{th}$=4 Post-$V_{th}$=3. The constant-fraction factor is appropriate to acquire time information for fast rise time and low noise signal (particular for signals from XP20D0 and LaBr$_3$ detector for 511 keV γ ray test) .The constant-fraction factor and the fitting range are two important facors for determining a timing moment with the minimum time jitter and eliminate the influence of the noise for accurate time measurement .The experiment results show that the time performance of the digital system based on the DRS4 module using the dCFD method is the best of several digital methods and the analog system investigated. Based on the processing time for the algorithms of each digital method, the CFD method has the fastest calculating speed.

The DRS4 module is inexpensive, has low power dissipation, high channel density and small size. Also, it does not require additional electronics for pulse height analysis and timing measurements. The dCFD method provides the best time resolution, which allows us to define the signal's leading edge and to get a time measurement point as close as possible to the baseline. The excellent performance demonstrates that the module, with its good time resolution (less than 200 ps), is suitable to better localize positron annihilation, improve the quality of medical images and reduce exposure times for TOF-PETs. We have already applied the DRS4 module in a PET system and obtained very clear PET crystal flood images and good position resolution, as reported in Ref. [14].

In the future, we will improve and apply the DRS4 to readout and acquisition signals for silicon photo-multipliers (SiPMs) in TOF-PET systems [15] and fast scintillator detectors in particle identification experiments [16] and use the dCFD method to process the raw data for better timing measurement with picosecond accuracy.

**Acknowledgements**

This work is supported by the Foundation of the Chinese Academy of Sciences (210340XBO),the National Natural Science Foundation of China (11305233,11205222) ,General Program of National Natural Science Foundation of China (11475234),Youth Innovation Promotion Association, CAS (201330YQO), Specific Fund of National key scientific instrument and equipment development project (2011YQ12009604) and Joint Fund for Research Based on Large-Scale Scientific Facilities (U1532131).